\documentclass[a4paper]{mem}
\usepackage{natbib}
\usepackage{graphicx}
\usepackage[a4paper]{hyperref}
\idline{1}{1}
\begin{document}
   \title{The role of nova nucleosynthesis in\\ Galactic chemical evolution
         }

   \author{Donatella Romano \inst{1}
          \and
          Francesca Matteucci \inst{2}
          }

   \offprints{D. Romano}
   \mail{via Ranzani 1, 40127 Bologna}

   \institute{INAF\,--\,Osservatorio Astronomico di Bologna,
             via Ranzani 1, 40127 Bologna \email{romano@bo.astro.it}\\ 
             \and
             Dipartimento di Astronomia, Universit\`a di Trieste,
	     via Tiepolo 11, 34131 Trieste \email{matteucci@ts.astro.it}
             }

   \abstract{In this paper we study the impact of nova nucleosynthesis on 
            models for the chemical evolution of the Galaxy. It is found that 
	    novae are likely to be the sources of non-negligible fractions of 
	    the $^7$Li and $^{13}$C observed in disc stars. Moreover, they 
	    might be responsible for the production of important amounts of 
	    $^{17}$O at late times and probably account for a major fraction 
	    of the Galactic $^{15}$N.
            }
   \authorrunning{D. Romano and F. Matteucci}
   \titlerunning{Nova nucleosynthesis and chemical evolution}
   \maketitle
%

\section{Introduction}

Classical novae are binary systems consisting of a CO or ONeMg white dwarf 
accreting hydrogen-rich matter from a main-sequence companion, which 
sporadically inject nuclearly processed material into the interstellar medium 
(ISM). The thermonuclear runaway (TNR), responsible for the explosion causing 
the ejection of almost the whole previously accreted envelope, also leads to 
the synthesis of some rare nuclei: $^7$Li, $^{13}$C, $^{15}$N, $^{17}$O, 
$^{22}$Na and $^{26}$Al (e.g., Starrfield et al. 1972, 1974, 1978; Politano et 
al. 1995; Hernanz et al. 1996;  Jos\'e \& Hernanz 1998). Here we address the 
issue of the evolution of $^7$Li and CNO isotopes in the Milky Way, putting 
emphasis on the contribution from nova systems.

\section{Lithium evolution in the solar neighborhood}

   \begin{figure*}
   \centering
   \includegraphics[width=6.5cm]{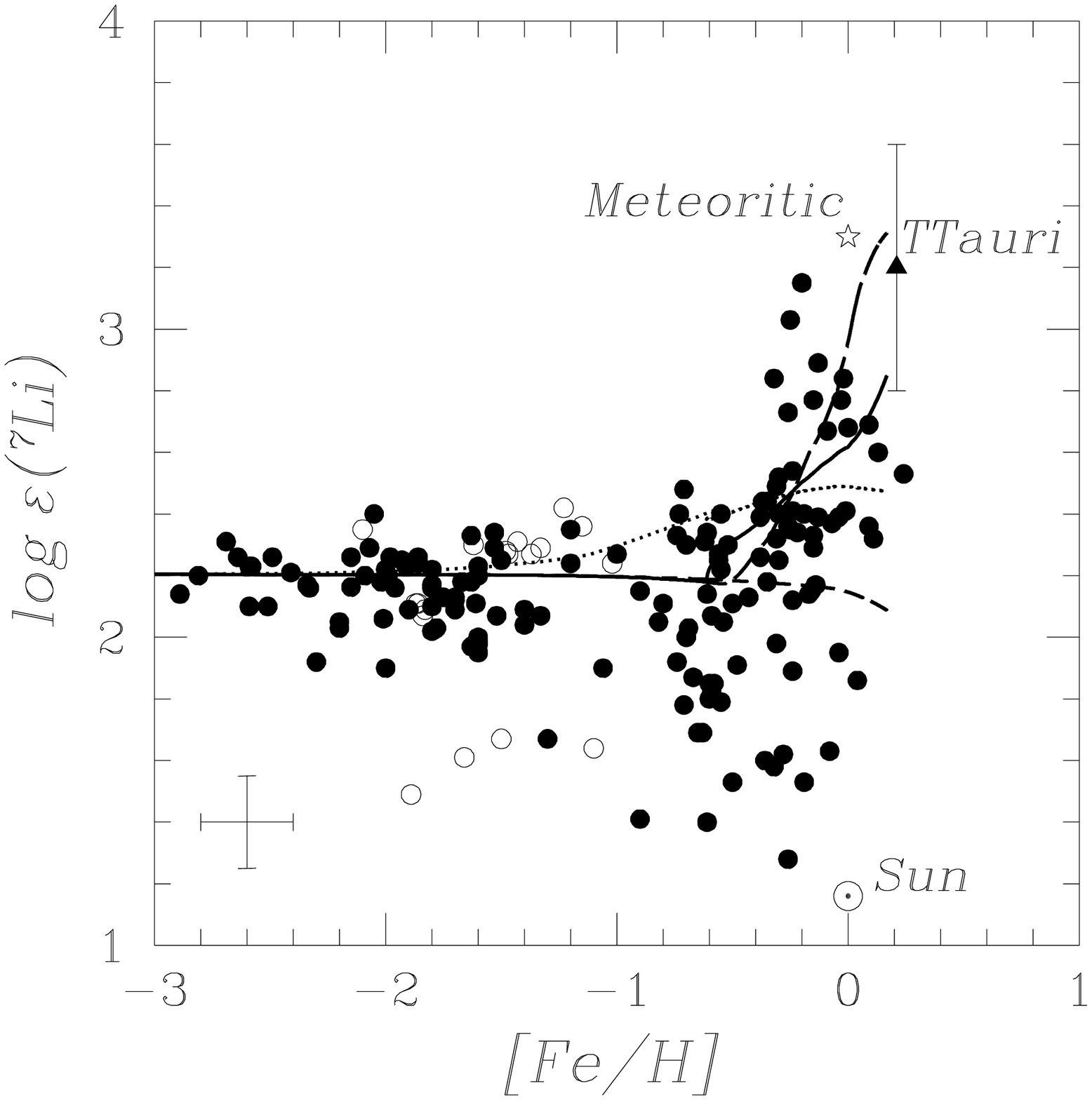}
   \includegraphics[width=6.5cm]{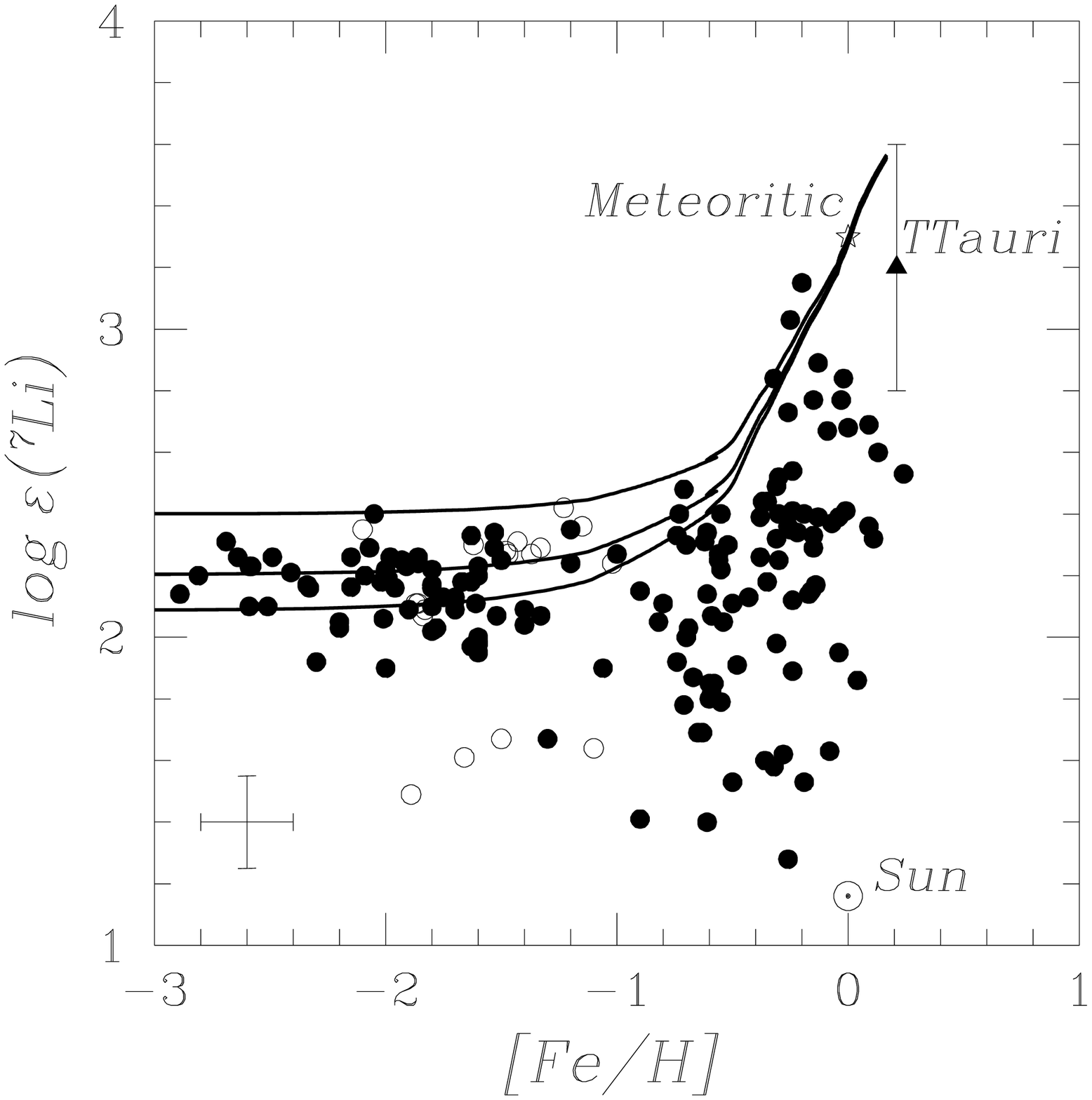}
   \caption{Evolution of $^7$Li in the solar vicinity. Data are from the 
           compilation of Romano et al. (1999) (dots) and from Ryan et al. 
	   (2001) (empty circles). The meteoritic, T\,Tauri and Sun values are 
           shown as well. Explanation of the different models is given in the 
	   text.}
   \label{figlit}
   \end{figure*}

The upper envelope of the observed $\log \varepsilon$($^7$Li) versus [Fe/H] 
diagram, as traced by warm ($T_{\mathrm{eff}} \ge 5700$ K) field dwarfs in the 
solar neighborhood (see Fig.~1), is generally believed to reflect the $^7$Li 
enrichment history of the ISM due to different production processes which rise 
its content from the primordial abundance to the higher values observed in 
meteorites and local ISM (e.g., D'Antona \& Matteucci 1991; Matteucci et al. 
1995; Romano et al. 1999, 2001). While halo dwarfs lying on the \emph{Spite 
plateau} share almost the same lithium abundance, a broad spread in the data 
is seen at high metallicities. This is commonly explained as due to lithium 
dilution/destruction mechanisms, which activate at high metallicities. In 
Fig.~1, left panel, we show the results of four chemical evolution models; in 
each of them $^7$Li is produced by a single category of stellar $^7$Li 
producers: asymptotic giant branch (AGB) stars (short-dashed line), Type II 
supernovae (SNeII) (dotted line), novae (solid line), and low-mass red giants 
(long-dashed line). Novae and low-mass stars on the red giant branch (RGB) are 
the best candidates in order to reproduce the observed rise off the plateau 
(see Romano et al. 1999, 2001 for details). In Fig.~1, right panel, we show 
the results of models where all the different Li sources of Fig.~1, left 
panel, are taken into account. Lithium production from Galactic cosmic rays 
(GCRs) is included as well. These models differ only in the adopted primordial 
Li abundance. It can be immediately seen that lithium evolution during most of 
the Galactic lifetime is practically independent of the adopted primordial 
$^7$Li abundance. Therefore, {\it we conclude that novae should contribute a 
non-negligible Li amount} independently of the assumed primordial Li 
abundance. This result is particularly relevant in view of recent{\it WMAP} 
data, suggesting a high primordial lithium abundance of $\log 
\varepsilon$($^7$Li)$_{\mathrm{p}} \sim 2.6$ (see Romano et al. 2003 for a 
discussion on this point).

\section{CNO isotope evolution in the Milky Way}

   \begin{figure*}
   \centering
   \includegraphics[width=13cm]{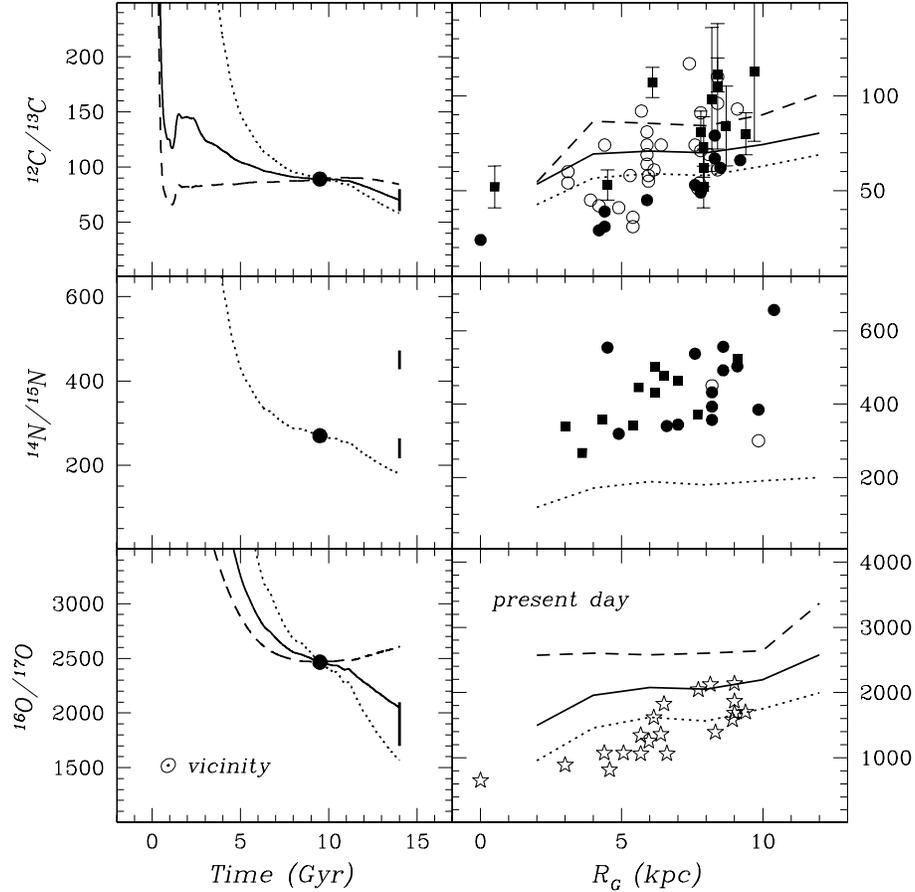}
   \caption{Theoretical evolution of $^{12}$C/$^{13}$C (upper panels), 
           $^{14}$N/$^{15}$N (middle panels) and $^{16}$O/$^{17}$O (lower 
           panels) in the solar vicinity (left panels) and across the Galactic 
	   disc at the present time (right panels), versus observations. See 
	   text for details.}
   \label{figcno}
   \end{figure*}

Among the CNO group nuclei, $^{16}$O is the best understood. It is a primary 
element, i.e., it is always synthesized starting from H and He in the parent 
star. The bulk of $^{16}$O comes from massive stars ($m \/ > 10 \; M_\odot$). 
The production of $^{12}$C is more uncertain: it is synthesized as a primary 
element, but the exact amount restored into the ISM by stars of different 
masses is still uncertain. We favor the hypothesis that $^{12}$C is mostly 
produced by low- and intermediate-mass stars (LIMS) (Chiappini et al. 2003). 
$^{14}$N should be a typical secondary element, but with a primary component 
as well, coming mostly from LIMS (Chiappini et al. 2003; Chiappini \& 
Matteucci, these proceedings).

The TNR leading to nova outbursts has been identified as a promising channel 
for the synthesis of $^{13}$C, $^{15}$N and $^{17}$O (see references quoted in 
the Introduction). $^{13}$C and possibly $^{17}$O are also formed in the 
external regions of stars in the RGB, planetary nebula and SN phase, while 
part of $^{15}$N production could be due to rotationally induced mixing of 
protons into the helium-burning shells of massive stars (Chin et al. 1999, and 
references therein). In the following, we try to ascertain whether the 
contribution from novae is needed in order to reproduce the observations 
concerning both the temporal and the spatial variation of the CNO isotopic 
ratios in the Galaxy.

The data on the CNO isotopic ratios in the Galaxy suggest that $^{13}$C, 
$^{15}$N and $^{17}$O behave as secondary elements. In fact, their abundances 
do increase with increasing time and decreasing Galactocentric distance, i.e., 
with metallicity (see Romano \& Matteucci 2003 for a summary of the available 
data).

Here we show results from:
\begin{itemize}
 \item a model considering CNO production only from single low-, intermediate- 
       and high-mass stars (Model 1);
 \item a model assuming novae as the sole producers of $^{13}$C, $^{15}$N and 
       $^{17}$O (Model 2);
 \item a model where $^{13}$C, $^{15}$N and $^{17}$O are produced both by 
       novae and single stars (Model 3).
\end{itemize}
The nucleosynthesis prescriptions are from van den Hoek \& Groenewegen (1997) 
and Ventura et al. (2002) for LIMS; Nomoto et al. (1997) for massive stars; 
Jos\'e \& Hernanz (1998) for novae. $^{13}$C, $^{15}$N and $^{17}$O from novae 
are assumed to be of primary origin in both Models 2 and 3. However, from the 
point of view of Galactic chemical evolution, they behave as secondary 
elements, owing to the long time-scales on which they are restored into the 
ISM by their long-lived stellar progenitors (details can be found in Romano \& 
Matteucci 2003).

In Fig.~2, left panels, we display the temporal evolution of the carbon (upper 
panel), nitrogen (middle panel) and oxygen (lower panel) isotopic ratios in 
the solar neighborhood as predicted by Models 1 (dashed line), 2 (dotted line) 
and 3 (solid line). The meteoritic and local values are shown as well (see 
Romano \& Matteucci 2003 for references). In Fig.~2, right panels, model 
predictions regarding the spatial variation along the disc at the present time 
are compared to the data (models are labeled according to Fig.~2, left panels).

{\it A model in which $^{\mathit{13}}$C and $^{\mathit{17}}$O are produced by 
intermediate- and high-mass stars as well as novae fits better the 
observations} regarding both the temporal and the spatial variation of the CNO 
isotopic ratios in the Milky Way; however, in the case of $^{17}$O we find 
that it is necessary to lower by hand the yields of both intermediate-mass 
stars and novae in order to reproduce the solar $^{17}$O abundance (Romano \& 
Matteucci 2003). The behaviour of the nitrogen isotopic ratio along the 
Galactic disc seems to suggest that $^{15}$N has to be produced on long 
time-scales, and {\it novae are the best candidates for producing 
$^{\mathit{15}}$N on long time-scales in the framework of the presently 
available nucleosynthesis calculations}. However, it is apparent that the 
issue of $^{15}$N nucleosynthesis in stars deserves further investigation.

\bibliographystyle{aa}

\begin{thebibliography}{}

\bibitem[{Chiappini et al. (2003)}]{chi03}
Chiappini, C., Romano, D., \& Matteucci, F. 2003, MNRAS 339, 63
\bibitem[{Chin et al. (1999)}]{chi99}
Chin, Y., Henkel, C., Langer, N., \& Mauersberger, R. 1999, ApJ 512, L143
\bibitem[{D'Antona \& Matteucci (1991)}]{danmat91}
D'Antona, F., \& Matteucci, F. 1991, A\&A 248, 62
\bibitem[{Hernanz et al. (1996)}]{her96}
Hernanz, M., Jos\'e, J., Coc, A., \& Isern, J. 1996, 
   ApJ 465, L27
\bibitem[{Jos\'e \& Hernanz (1999)}]{josher98}
Jos\'e, J., \& Hernanz, M. 1998, ApJ 494, 680
\bibitem[{Matteucci et al. (1995)}]{mat95}
Matteucci, F., D'Antona, F., \& Timmes, F. X. 1995, A\&A 303, 460
\bibitem[{Nomoto et al. (1997)}]{nom97}
Nomoto, K., Hashimoto, M., Tsujimoto, T., Thielemann, F.-K., Kishimoto, N., 
   Kubo, Y., \& Nakasato, N. 1997, Nucl. Phys. A 616, 79c
\bibitem[{Politano et al. (1995)}]{pol95}
Politano, M., Starrfield, S., Truran, J. W., Weiss, A., \& Sparks, W. M. 1995, 
   ApJ 448, 807
\bibitem[{Romano \& Matteucci (2003)}]{rommat03}
Romano, D., \& Matteucci, F. 2003, MNRAS 342, 185
\bibitem[{Romano et al. (1999)}]{rom99}
Romano, D., Matteucci, F., Molaro, P., \& Bonifacio, P. 1999, A\&A 352, 117
\bibitem[{Romano et al. (2001)}]{rom01}
Romano, D., Matteucci, F., Ventura, P. \& D'Antona, F. 2001, A\&A 374, 646
\bibitem[{Romano et al. (2003)}]{rom03}
Romano, D., Tosi, M., Matteucci, F., \& Chiappini, C. 2003, MNRAS submitted
\bibitem[{Ryan et al. (2001)}]{ryan01}
Ryan, S. G., Beers, T. C., Kajino, T., \& Rosolankova, K. 2001, ApJ, 547, 231
\bibitem[{Starrfield et al. (1972)}]{star72}
Starrfield, S., Truran, J. W., Sparks, W. M., \& Kutter, G. S. 1972, 
   ApJ 176, 169
\bibitem[{Starrfield et al. (1974)}]{star74}
Starrfield, S., Sparks, W. M., \& Truran, J. W. 1974, 
   ApJ 192, 647
\bibitem[{Starrfield et al. (1978)}]{star78}
Starrfield, S., Truran, J. W., Sparks, W. M., \& Arnould, M. 1978, 
   ApJ 222, 600
\bibitem[{van den Hoek \& Groenewegen (1997)}]{vangro97}
van den Hoek, L. B., \& Groenewegen, M. A. T. 1997, A\&AS 123, 305
\bibitem[{Ventura et al. (2002)}]{ven02}
Ventura, P., D'Antona, F., \& Mazzitelli, I. 2002, A\&A 393, 215

\end{thebibliography}

\end{document}